%% file: main.tex
\documentclass[sigconf,authorversion,nonacm]{acmart}

\AtBeginDocument{%
  \providecommand\BibTeX{{%
    \normalfont B\kern-0.5em{\scshape i\kern-0.25em b}\kern-0.8em\TeX}}}

\usepackage{booktabs} 
\usepackage{amsmath, amsfonts}
\usepackage[ruled,vlined]{algorithm2e}
\usepackage{algpseudocode}
\usepackage{graphicx}
\usepackage{multirow}
\usepackage{multicol}
\usepackage{subcaption}
\usepackage{textcomp}
\usepackage{xcolor}
\usepackage{diagbox}
\usepackage[]{hyperref}
\usepackage{enumitem}
\setlist[itemize]{leftmargin=*}
\usepackage{url}
\usepackage{threeparttable}
\usepackage{pifont}
\usepackage{nopageno}
\usepackage{bm}
\usepackage{scalerel}
\usepackage{commath}
\usepackage{tikz}
\usepackage{soul}
\usepackage{lipsum}
\usepackage{mathtools}
\usepackage{cuted}
\usepackage{stfloats}

\newcommand{\R}{\mathbb{R}}
\makeatletter
\renewcommand\footnoterule{
  \kern-3\p@
  \hrule\@width\columnwidth
  \kern2.6\p@}
  \makeatother

\makeatletter
\newcommand\footnoteref[1]{\protected@xdef\@thefnmark{\ref{#1}}\@footnotemark}
\makeatother

\newcommand\scalemath[2]{\scalebox{#1}{\mbox{\ensuremath{\displaystyle #2}}}}
\algnewcommand\algorithmicforeach{\textbf{for each}}
\algdef{S}[FOR]{ForEach}[1]{\algorithmicforeach\ #1\ \algorithmicdo}

\setcopyright{none}
\begin{document}

\input{CCS_concept}


\title{A Fast Semi-Analytical Approach for Transient Electromigration Analysis of Interconnect Trees using Matrix Exponential}

\fancyhead{} 
\fancyfoot{}

\author{Pavlos~Stoikos}
 \affiliation{
  \institution{Dept. of Electrical \& Computer Eng.}
  \city{University of Thessaly, Volos}
  \country{Greece}
}
\email{pastoikos@e-ce.uth.gr}

\author{George~Floros}
 \affiliation{
  \institution{Dept. of Electrical \& Computer Eng.}
  \city{University of Thessaly, Volos}
  \country{Greece}
}
\email{gefloros@e-ce.uth.gr}

\author{Dimitrios~Garyfallou}
 \affiliation{
  \institution{Dept. of Electrical \& Computer Eng.}
  \city{University of Thessaly, Volos}
  \country{Greece}
}
\email{digaryfa@e-ce.uth.gr}

\author{Nestor~Evmorfopoulos}
 \affiliation{
  \institution{Dept. of Electrical \& Computer Eng.}
  \city{University of Thessaly, Volos}
  \country{Greece}
}
\email{nestevmo@e-ce.uth.gr}

\author{George~Stamoulis}
 \affiliation{
  \institution{Dept. of Electrical \& Computer Eng.}
  \city{University of Thessaly, Volos}
  \country{Greece}
}
\email{georges@e-ce.uth.gr}

\input{abstract}
\maketitle

\input{1_Intro}
\input{2_Background}
\input{3_Proposed}

\input{4_Results}
\input{5_Conclusions}
\input{acknowledgement}

\bibliography{refs}
\bibliographystyle{tran}

\end{document}

%% file: CCS_concept.tex
\ccsdesc[300]{General and reference~Reliability}
\ccsdesc[300]{Hardware~Metallic interconnect}
\ccsdesc[300]{Hardware~Aging of circuits and systems}
\ccsdesc[300]{Hardware~Power grid design}


%% file: abstract.tex
\begin{abstract}
As integrated circuit technologies are moving to smaller
technology nodes, Electromigration (EM) has become one of the most challenging problems facing the EDA industry. While numerical approaches have been widely deployed since they can handle complicated interconnect structures, they tend to be much slower than analytical approaches. In this paper, we present a fast semi-analytical approach, based on the matrix exponential, for the solution of Korhonen's stress equation at discrete spatial points of interconnect trees, which enables the analytical calculation of EM stress at any time and point independently. The proposed approach is combined with the extended Krylov subspace method to accurately simulate large EM models and accelerate the calculation of the final solution. Experimental evaluation on OpenROAD benchmarks demonstrates that our method achieves 0.5\% average relative error over the~COMSOL industrial tool while being up to three orders of magnitude~faster. 
\end{abstract}

%% file: 1_Intro.tex
\section{Introduction}
\label{sec:intro}
Electromigration (EM) has become one of the greatest concerns for the semiconductor industry in recent years.  EM failures 
constitute an inevitable consequence
of the rising current demands and the smaller process geometries, and may lead to a number of open- or short-circuits in on-chip interconnects \cite{yang_2017}. To this end, EM analysis has become 
an
integral part of modern VLSI design~flows~\cite{lienig_book_2018}.

In the past, several empirical methods for EM analysis have been developed, such as the application of the Blech criterion \cite{blech_1976} followed by the Black's equation \cite{black_1969}. These two methods are generally applied together to identify  potential ``immortal'' wires and then predict the mean time to failure for the rest of them. However, besides the heuristic nature of these approaches, which have become inaccurate for modern technology nodes \cite{sachin_ispd2019}, these methods are based on single-segment wire structures, while modern VLSI interconnects contain multiple trees that are continuously connected forming complex wire structures.

Contrary to the previous empirical approaches, Korhonen et al. \cite{8de255f92cd646d2a40644644dc0d089} formed an exact physics-based model as
diffusion-like Partial Differential Equations (PDEs). Building on this work, several methodologies have been developed for computing the EM stress in segment lines. More specifically, these methods can be divided into two main categories. 
First, numerical methods, such as \cite{cook2018, smacd2022},  are well-established due to their simplicity and have already been integrated into commercial tools such as COMSOL~\cite{COMSOL}.
These methods perform discretization of space and time, and are in principle applicable to a wide spectrum of geometries due to the spatial discretization. However, they do not scale well and are computationally prohibitive for large-scale interconnects.

As a result,
the emphasis has been placed to 
analytical methods for the solutions of Korhonen's equations, which keep both space and time continuous, and can be effectively applied to large-scale systems. Previous analytical approaches \cite{chen-semi-analytical} calculate infinite series solutions and can be applied in general multi-segment~interconnects. Moreover, in \cite{iccad21}, the concept of stress reflections  was~introduced, which can also be applied to general multi-segment lines of arbitrary number of segments. However, the majority of these methods involve approximation of infinite series with a finite number of terms, which can become expensive since the number is dependent on both line length and time and cannot be known~beforehand.

In this paper, we present a fast semi-analytical approach for the solution of the Korhonen’s equation for general interconnect trees, which discretizes only space while keeping time continuous. The~main contributions of this paper are summarized hereafter. 
{\em First}, our method can calculate the EM stress for any given input time, by directly computing the analytical solution through the matrix exponential at any given point. 
{\em Second}, we develop a procedure for applying the Extended Krylov Subspace (EKS) in order to approximate the matrix exponential, which can significantly reduce the complexity of the proposed methodology. For the EM stress equation, we leverage that each segment in an interconnect structure is assumed to carry a constant current density \cite{7827687}. As a result, the subspace calculation is an one-time cost.
We evaluate our methodology on  available large-scale OpenROAD benchmarks and several artificial interconnect trees in order to prove the scalability of our method, while its efficiency and accuracy are validated against COMSOL by achieving great speedups and negligible error.

The rest of this paper is organized as follows. Section \ref{sec:background} provides basic background on EM analysis of interconnect trees. Section \ref{sec:problem_formulation} demonstrates the problem formulation of EM analysis.
Then, in Section \ref{sec:analytical_solution}, we present our main contributions in the analytical solution of the Korhonen's diffusion equation using the matrix exponential. Section \ref{sec:experimentals} demonstrates the experimental evaluation~of our method on available OpenROAD benchmarks and several artificial interconnect trees, followed by the conclusions in Section~\ref{sec:conclusions}.

%% file: 2_Background.tex
\section{Background}
\label{sec:background}

\subsection{Electromigration basics} 
\label{sec:EM_mechanism}
\begin{figure}[!hbt]
    \begin{center}
    \includegraphics[width=0.95\linewidth,keepaspectratio]{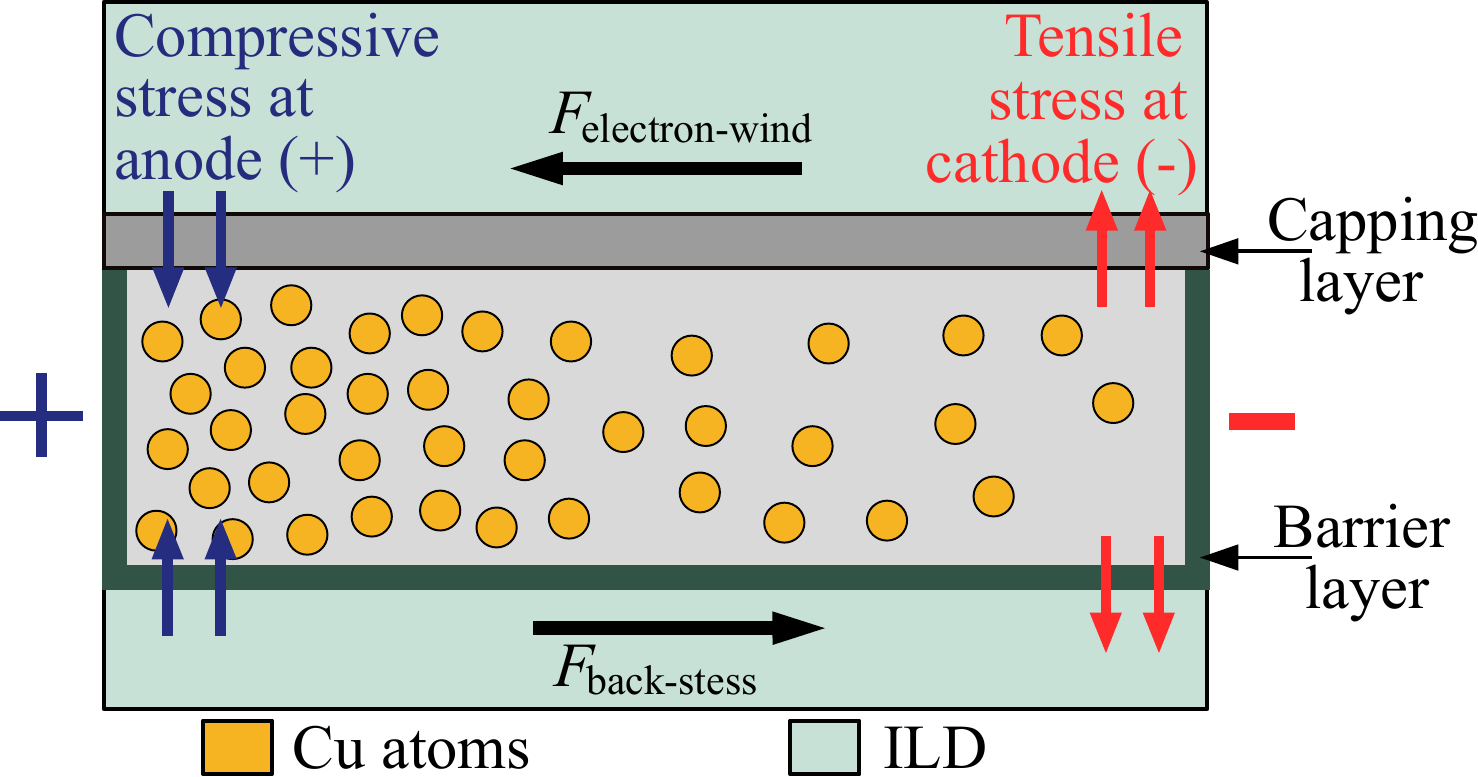}
     \end{center}
     \vspace{-8pt}
    \caption{Cross section of a Cu wire indicating the two forces. 
    }
    \label{fig:1}
\end{figure}

As shown in Fig. \ref{fig:1}, which is the cross section of a Cu Dual Damascene (DD) wire, the movement of  metal atoms is mainly determined by the resultant of two opposing forces. The first one, the $F_{\text{electron-wind}}$ is generated by the momentum transfer  between electrons and metal atoms, and is the primary cause of EM. The second one, known as $F_{\text{back-stress}}$, is an electrostatic force caused by the electric field strength in the metal atoms and has a direction opposite to the electron flow. Since  $F_{\text{back-stress}}$ is negligible compared to $F_{\text{electron-wind}}$ \cite{SULLIVAN1967347}, the movement of metal atoms occurs in the direction of the current flow, from the cathode (-) to the anode (+). As time passes, the disparity in the concentration of metal atoms between the anode and the cathode creates a compressive stress in the former and a tensile stress  in the latter. This causes a hillock formation near the anode and a void formation at the cathode, leading to open circuits and the end of the wire's lifetime.

It is well known that the movement of  metal atoms in a Cu DD interconnect technology is limited to one layer due to the diffusion barriers, preventing the mass transport to adjacent layers
\cite{5510581}.
Consequently, EM analysis of a huge interconnect structure, such as a power grid, can be performed layer by layer.  
More specifically, each layer has a generic orthogonal mesh structure that can be divided into a group of multi-segment interconnect trees \cite{6549125}.
Based on that, analysis of every multi-segment interconnect tree is carried out, on which this paper is mainly focused.

\begin{figure}[!hbt]
    \begin{center}
    \includegraphics[width=1\linewidth,keepaspectratio]{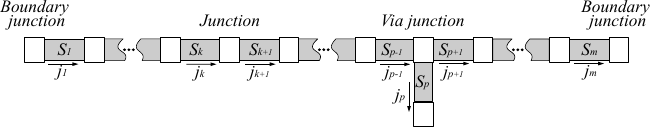}
    \end{center}
    \vspace{-8pt}
    \caption{An \textit{m}-segment interconnect tree.}
    \label{fig:general_tree}
\end{figure}

Generally, in a multi-segment structure, a set of wire segments and vias are interconnected with junctions, where each wire segment carries a certain current
density. Fig. \ref{fig:general_tree} depicts an example of a multi-segment interconnect tree \cite{iccad21}.

\subsection{Korhonen's model}
\label{sec:korhonen}
    According to the EM analysis of a multi-segment interconnect tree, the stress evolution  $\sigma(x,t)$ of each segment can be described by the diffusion  Korhonen's  PDE \cite{8de255f92cd646d2a40644644dc0d089}, which relates the stress $\sigma$ to the distance from the cathode $x$. Their relation is formed as:
\begin{equation}
\label{korhonen}
\frac{\partial \sigma}{\partial t} = \frac{\partial}{\partial x}\left[\kappa\left(\frac{\partial\sigma}{\partial x}+ \beta j\right)\right]
\end{equation}

\noindent
, where $\beta = (Z^{*}e\rho)/\Omega$ is the EM driving force and $\kappa = D_{a}\mathcal{B}\Omega/ (k_{B}T)$ is the diffusivity of stress with $D_{a} = D_{0}e^{-E_{a}/k_{B}T}$ being the diffusion coefficient. Here, $E_{a}$ is the activation energy, $D_{0}$ is the diffusivity constant,  $j$ is the current density through the segment of the wire, $Z^{*}$ is the effective charge number, $e$ is the electron charge, $\rho$ is the resistivity, $\Omega$ is the atomic volume for the metal, $\mathcal{B}$ is the bulk modulus of the material, $k_{B}$ is Boltzmann's constant, $T$ is the temperature, $x$ is the coordinate along the wire, and $t$ is time. In addition, the stress gradient $\partial\sigma/ \partial x$ accounts the flux related to  $F_{\text{back-stress}}$, the term $\beta j$ represents atomic flux attributable to  $F_{\text{electron-wind}}$, while the sum of these two $\left(\partial\sigma /\partial x+ \beta j\right)$ is   related to the net atomic flux.

The described Eq. (\ref{korhonen}) is supplemented by a  set of  boundary conditions that relate the stress at any point $x_{i}$, along with a temporal boundary condition that initializes the stress values at $ t = 0$ \cite{7827687}. For an intermediate point $x_{i}$ of the structure with degree $d_{i}$, the set of incident segments is denoted as $S_{i} = \{s_{1},s_{2}, \dots ,s_{d_{i}}\}$. Therefore, the spatial boundary conditions are as follows:
\\
$\mathbf{(1)}$ $\mathbf{Continuity}$ $\mathbf{constraints}$:  At any intermediate point $x_{i}$ of the multi-segment tree, the stress must be continuous:
\begin{equation}
\label{bc1}
\sigma_{s_{k}}(x=x_{i},t) = \sigma_{s_{k+1}}(x=x_{i},t),  \quad k = 1,\dots ,d_{i}-1
\end{equation}
\noindent
\\
$\mathbf{(2)}$ $\mathbf{Flux}$ $\mathbf{constraints}$:  The total atomic flux entering each point $x_{i}$ must be equal to zero:
\begin{equation}
\label{bc2}
\sum_{s_{k} \in S_{i}}^{} 
w_{s_{k}}\kappa_{s_{k}}\left( \frac{\partial\sigma_{s_{k}}}{\partial x}\Big|_{x=x_{i}}+ \beta j_{s_{k}} \right) = 0
\end{equation}

\noindent, where $\kappa_{s_{k}}$ is the diffusivity of segment $s_{k}$, $j_{s_{k}}$ is the current density of segment $s_{k}$, which is positive  when directed away from $x_{i}$ and negative when directed into $x_{i}$, and $w_{s_{k}}$ is the segment width. 
A special case is that the boundary conditions at any end-point (i.e., point with degree 1) require zero
flux across the blocking boundary,~i.e., 
\begin{equation}
\label{bc3}
 \frac{\partial\sigma_{s_{k}}}{\partial x}\Big|_{x=x_{i}}+ \beta j_{s_{k}}  = 0
\end{equation}

\noindent 
, where  $j_{s_{k}}$ is the current density of segment $s_{k}$ that is incident on end-point $x_{i}$. 

\section{Problem  Formulation}
\label{sec:problem_formulation}

In EM analysis, each segment of the interconnect tree is considered to carry a constant current density $j$ \cite{7827687}. As a result, the Korhonen's Eq. (\ref{korhonen})  for each segment takes the following form:
\begin{equation}
\label{korhonen_simple}
\frac{\partial \sigma}{\partial t} = \kappa\frac{\partial^{2}\sigma}{\partial x^{2}}
\end{equation}

\begin{figure}[!htb]
    \begin{center}
    \includegraphics[width=1\linewidth,keepaspectratio]{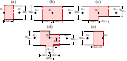}
    \end{center}
        \vspace{-8pt}
         \caption{
          Each discretized point may be located at (a) the left boundary segment, (b) the middle of  a segment, (c) an intermediate junction, (d) a via junction, and (e) the right boundary segment.The red box around each discretized point represents the corresponding area of Eq.~(\ref{area_eq}).
         }
    \label{fig:area}
\end{figure}

This equation can be uniformly discretized  by dividing each segment of the interconnect tree into points of equal length $\Delta x$, using the Finite Difference Method (FDM). By applying a finite difference approximation of the spatial derivative in Eq. (\ref{korhonen_simple}), for each discretized point $i$ of the interconnect tree, we get:
\begin{equation}
\label{korhonen_simple_desc}
\frac{d \sigma_{i}}{d t} = \kappa\frac{(\frac{\sigma_{i+1} - \sigma_{i}}{\Delta x}) - (\frac{\sigma_{i} - \sigma_{i-1}}{\Delta x})}{\Delta x}
\end{equation}
i.e.,
\begin{equation}
\label{korhonen_simple_desc_w}
 (w  \cdot \Delta x )\frac{d \sigma_{i}}{d t} = \kappa\left(\frac{\sigma_{i+1} - \sigma_{i}}{\Delta x/w}\right) - \left(\frac{\sigma_{i} - \sigma_{i-1}}{\Delta x/w}\right)
\end{equation}

\noindent
, where $w$ is the width of the wire segment that point $i$  belongs to and $a_{i} = w \cdot \Delta x$ is the total area of the segments connected to the point $i$ (see Fig. \ref{fig:area}). The formula of $a_{i}$ is described as:
\begin{equation}
\label{area_eq}
a_{i} =\frac{1}{2}
\sum_{s_{k} \in S_{i}}^{} w_{s_{k}} \cdot \Delta x
\end{equation}
\noindent 
For the sake of simplicity, we consider thickness to be negligible compared to the other two dimensions and therefore can ignore~it.

After applying FDM on the $m$-segment interconnect tree of Fig.\ref{fig:general_tree}, the resulting $n$ discretized points may be located at five different locations, as shown in Fig. \ref{fig:area}. Considering the boundary conditions of Eq. (\ref{bc1}), (\ref{bc2}), and (\ref{bc3}), we can rewrite Eq. (\ref{korhonen_simple_desc_w}) for the $n$ discretized points into the Ordinary Differential Equation (ODE) system of Eq. (\ref{ode}). More specifically, Eq. (\ref{ode}) depicts the stamps  of the two boundary  points $x_{1}$ and $x_{n}$,  any point $x_{i}$ at the middle of a segment, any intermediate junction point $x_{i}$, and any via junction point $x_{j}$, with $1<i<j<n$.
As a result, we can write the ODE system~for~EM stress evolution as the following  Linear time-invariant (LTI) system:

\begin{figure*}[!t]
\begin{equation}
\begin{aligned}
\label{ode}
\scalemath{0.58}{
\left[ \begin{array}{lcccccr}
a_{1} & \cdots & 0 & \cdots & 0 &\cdots & 0 \\
\vdots & \ddots & \vdots & \ddots & \vdots & \ddots & \vdots\\
0 & \cdots & a_{i} & \cdots & 0 &\cdots & 0 \\
\vdots & \ddots & \vdots & \ddots & \vdots & \ddots & \vdots\\
0 & \cdots & 0 & \cdots & a_{j} &\cdots & 0 \\
\vdots & \ddots & \vdots & \ddots & \vdots & \ddots 
& \vdots\\
0 & \cdots & 0 & \cdots & 0 &\cdots & a_{n} 
 \end{array} \right]
\left[ \begin{array}{c}
\dot{\sigma_{1}}\\
\vdots\\
\dot{\sigma_{i}}\\
\vdots\\
\dot{\sigma_{j}}\\
\vdots\\
\dot{\sigma_{n}}
\end{array} \right]
}=
\scalemath{0.65}{
\frac{\kappa}{(\Delta x)}
\left[\begin{array}{lcccccccccr}
-w_{s_{1}} & w_{s_{1}} & 0 & \cdots & \cdots & \cdots  & \cdots & 0 \\
\vdots  & \ddots & \ddots & \ddots & \ddots & \ddots & \ddots & \vdots \\
0 & \cdots  &  w_{s_{k}} & - ( w_{s_{k}} +  w_{s_{k+1}}) &  w_{s_{k+1}} &\cdots &\cdots & 0 \\
\vdots & \ddots & \ddots & \ddots & \ddots & \ddots & \ddots & \vdots\\
0 & \cdots  & w_{s_{p-1}} & - ( w_{s_{p-1}} +  w_{s_{p}}  +  w_{s_{p+1}}) & w_{s_{p}} & w_{s_{p+1}} &\cdots & 0 \\
\vdots & \ddots & \ddots & \ddots & \ddots & \ddots & \ddots & \vdots\\
 0 & \cdots  & \cdots  & \cdots & \cdots & 0 & w_{s_{m}} & -w_{s_{m}}  \\

\end{array} \right]
\left[ \begin{array}{c}
\sigma_{1}\\
\sigma_{2}\\
\vdots\\
\sigma_{i-1}\\
\sigma_{i}\\
\sigma_{i+1}\\
\vdots\\
\sigma_{j-1}\\
\sigma_{j}\\
\sigma_{j+1}\\
\sigma_{j+2}\\
\vdots\\
\sigma_{n-1}\\
\sigma_{n}
\end{array} \right] + } 
\scalemath{0.65}{
\kappa\beta 
 \left[ \begin{array}{lcccccr}
w_{s_{1}} & 0 & \cdots & \cdots & \cdots& \cdots & 0 \\
\vdots & \ddots & \ddots & \ddots & \ddots & \ddots & \vdots  \\
0 &  \cdots &w_{s_{k}} &-w_{s_{k+1}} & \cdots & \cdots  & 0\\
\vdots & \ddots & \ddots & \ddots & \ddots & \ddots & \vdots \\
0 &  \cdots &-w_{s_{p-1}} &w_{s_{p}} &w_{s_{p+1}} &\cdots & 0\\
\vdots & \ddots & \ddots & \ddots & \ddots & \ddots & \vdots  \\
0 & \cdots & \cdots & \cdots & \cdots & 0  & -w_{s_{m}} \\
\end{array} \right]
\left[ \begin{array}{c}
j_{1}\\
\vdots\\
j_{k}\\
j_{k+1}\\
\vdots\\
j_{p-1}\\
j_{p}\\
j_{p+1}\\
\vdots\\
j_{m}

\end{array} \right]}
\end{aligned}
\end{equation}
\end{figure*}
\begin{equation}
\label{ode_system}
\mathbf{C}\dot{\mathbf{\sigma}}(t) = \mathbf{G} \mathbf{\sigma}(t) + \mathbf{B}\mathbf{j}(t) 
\end{equation}

%% file: 3_Proposed.tex
\section{Proposed  em stress analysis}
\label{sec:analytical_solution}

\subsection{Analytical solution }
\label{sec:mexp_background}
Initially, we elaborate on the matrix exponential time integration that describes the EM  stress evolution. 
The solution of~Eq.~(\ref{ode_system}) can be obtained analytically \cite{10.5555/578731}. First,  Eq. (\ref{ode_system}) can be written~as:
\begin{equation}
\dot{\mathbf{\sigma}}(t) = \mathbf{A} \mathbf{\sigma}(t) + \mathbf{b}(t) 
\end{equation}
with 
\begin{equation}
\mathbf{A} = \mathbf{C}^{-1}\mathbf{G}, \quad \mathbf{b}(t) = \mathbf{C}^{-1}\mathbf{B}\mathbf{j}(t)
\end{equation}

\noindent
Given that the stress values for all points at $t = 0$ are known~\cite{7827687}, the solution at $t$ can be obtained by:
\begin{equation}
\label{sol_integral}
\mathbf{\sigma}(t) = e^{t\mathbf{A}}\mathbf{\sigma}(0) + \int_{0}^{t} e^{(t-\tau)\mathbf{A}}\mathbf{b}(\tau) \,d\tau 
\end{equation}

\noindent
Since the input vector $\mathbf{j}(t)$ is constant, as described in Section \ref{sec:problem_formulation},  we can integrate the last term of Eq. (\ref{sol_integral}) analytically, transforming the solution to:
\begin{equation}
\begin{aligned} 
\label{finstate}
\mathbf{\sigma}(t) = e^{t\mathbf{A}} \left( \mathbf{\sigma}(0) + \mathbf{F}(t)  \right) -  \mathbf{F}(t)
\end{aligned}
\end{equation} \\
with
\begin{equation}
\label{eq_ft}
\mathbf{F}(t) = \mathbf{A}^{-1}\mathbf{b}(t)
\end{equation}
\noindent

\noindent

While calculating the analytical solution, two major drawbacks arise. The first is the singularity of the ODE LTI system.
More specifically, matrix $\mathbf{G}$ is non-invertible, 
rendering the calculation of the term $\mathbf{F}(t)$ impossible. 
This occurs due to the fact that the stress equations for discretized points are not independent, as there is no "ground" stress node.
The second drawback is the significant computational and memory cost of the matrix exponential operator, which stems from the increased order $n$ of the LTI system for finer discretizations or larger trees.

\subsection{Singularity elimination for the LTI system}
\label{sec:proposed_singularity}
To resolve the singularity of the LTI system, we need to apply an additional independent equation.
According to Korhonen's equations, the incoming and outcoming mass transport rates balance out for every discretized point.
As a result, the LTI sytem can be extended with the following independent equation:
 \begin{equation}
 \label{mass_conv}
\sum_{1}^{n} a_{i}\sigma_{i} = 0
 \end{equation}
 
\noindent
, where $a_{i}$ can be calculated using Eq. (\ref{area_eq}).
This equation describes the mass conservation in the stress kinetics and can be used  to substitute a dependent row of  matrix $\mathbf{G}$.
The singularity elimination process is presented in Algorithm \ref{elm_algo}. As can be seen, the substitution of any $k$-th row of matrix $\mathbf{G}$ using  Eq. (\ref{mass_conv}) is performed in step 1, while the $k$-th row of matrix $\mathbf{G}$ is eliminated  in steps 2-11.  

\LinesNumbered

\begin{algorithm}[hbt!]
\SetAlgoLined
\DontPrintSemicolon
\KwIn{ singular matrix $\mathbf{G}$, equation $\mathbf{Eq}_{\text{mass}}$ of Eq. (\ref{mass_conv}),  original order $\mathbf{n}$, point $x_{k}$, set  $\mathbf{N}_{k}$ of points incident on $x_{k}$ }
\KwOut{ nonsingular matrix $\mathbf{G}$ }

 $\mathbf{G}{(k,:)} = \mathbf{Eq}_{\text{mass}}$\\
 \ForEach{ point $ \upsilon \in N_{k}$}{%
     \For{$j=1,\dots,n$}{
        \If{ $ j \neq k $ }{
          $\mathbf{w} = \mathbf{G}(\upsilon, k)\mathbf{G}^{-1}(k,k)\mathbf{G}(\upsilon, j)$\\
          $\mathbf{G}(\upsilon, j) =\mathbf{G}(\upsilon, j) - \mathbf{w}$\\
        }
     }
 }
 
$\mathbf{G}^{'}=[\mathbf{G}(1:k-1,:), \mathbf{G}(k+1:N,:)]$\\
 $\mathbf{G}^{}=[\mathbf{G}^{'}(:,1:k-1), \mathbf{G}^{'}(:,k+1:N)]$\\
 
\caption{Singularity elimination for the LTI system}%
\label{elm_algo}
\end{algorithm}
\normalsize

\subsection{Krylov subspace-based EM stress analysis}
\label{sec:proposed_eks}
For efficiently computing the matrix exponential, we present a model order reduction approach based on Arnoldi procedure that estimates the stress values of a multi-segment interconnect tree for every given time $t$. 
On one hand, the Arnoldi process produces the Hesseberg matrix $\mathbf{H}_{m}$ ($m\ll n$)  of a Krylov subspace that tends to approximate the large magnitude eigenvalues of  matrix $\mathbf{A}$.
On the other hand, EM is a very low process whose dynamics are dominated by the lowest magnitude eigenvalues of  matrix $\mathbf{A}.$
Therefore, the main idea is to apply the EKS~\cite{ASPDAC21}, in order to get the reduced order model approximating both the largest and the smallest eigenvalues of the original LTI system.
 The EKS is effectively the combination of the standard Krylov subspace $\mathcal{K}_{m/2}(\mathbf{A}_{E},\mathbf{B}_{E}) $ and the inverted subspace $\mathcal{K}_{m/2}(\mathbf{A}_{E}^{-1},\mathbf{B}_{E})$, i.e.,
 \begin{equation*}
\label{Eq:eksm}
\mathcal{K}_m^E(\mathbf{A}_{E},\mathbf{B}_{E}) = \mathcal{K}_{m/2}(\mathbf{A}_{E},\mathbf{B}_{E}) + \mathcal{K}_{m/2}(\mathbf{A}_{E}^{-1},\mathbf{B}_{E}) =
\end{equation*} 
\begin{equation}
\label{eks_eqn}
\begin{aligned} 
span \{\mathbf{B}_{E},  \mathbf{A}_{E}^{-1}\mathbf{B}_{E}, \mathbf{A}_{E}\mathbf{B}_{E},\mathbf{A}_{E}^{-2}\mathbf{B}_{E}, \mathbf{A}_{E}^{2}\mathbf{B}_{E},\mathbf{B}_{E},\dots,\\\mathbf{A}_{E}^{(m/2)-1}\mathbf{B}_{E},
\mathbf{A}_{E}^{-m/2}\mathbf{B}_{E}\}  
\end{aligned}
\end{equation} \\
, where:
\begin{equation}
\mathbf{A}_{E} \equiv \mathbf{G}^{-1}\mathbf{C},\quad \mathbf{B}_{E} \equiv  \mathbf{\sigma}(0) + \mathbf{F}(t) 
\end{equation}
\noindent
Note that, if the desired order is an odd number, the EKS subspace would be similar to Eq. (\ref{eks_eqn}) without the last column $\mathbf{A}_{E}^{-m/2}\mathbf{B}_{E}$.
The orthogonal basis of the EKS span is stored as the columns of a projection matrix $\mathbf{V}_{m} = [\boldsymbol{\upsilon}_{1}, \dots, \boldsymbol{\upsilon}_{m}] \in \R^{n \times m}$, which satisfies the so-called Arnoldi decomposition:
  \begin{equation}
  \label{decomp}
 \mathbf{A} \mathbf{V}_{m} =  \mathbf{V}_{m+1} \mathbf{H}_{m+1} =  \mathbf{V}_{m}\mathbf{H}_{m} + \mathbf{h}_{m+1,m}\boldsymbol{\upsilon}_{m+1}\mathbf{e}_{m}^{\intercal}
 \end{equation} 

\noindent
, where $\mathbf{H}_{m}  \in \R^{m \times m}$  is the upper Hessenberg matrix,  
which is the matrix $\mathbf{H}_{m+1}$ without the last row  $(0, \dots, 0,  \mathbf{h}_{m+1,m})$, and $\mathbf{e}_{m}^{\intercal} = [0, \dots, 0, 1]^{\intercal} \in \R^{m} $ is the last canonical basis vector in $\R^{m}$. Once $\mathbf{V}_{m}$ and $\mathbf{H}_{m}$ are generated,  an approximation $\boldsymbol{\phi}_{m}(t)$ to the matrix exponential $\boldsymbol{\phi}(t) = e^{t\mathbf{A}}\mathbf{v}$ is usually computed as:
 \begin{equation}
 \label{smapp}
\boldsymbol{\phi}_{m}(t) =  \beta\mathbf{V}_{m}\mathbf{H}_{m} e^{t\mathbf{H}_{m}}\mathbf{e}_{1} 
 \end{equation} 
 
 \noindent
, where $\mathbf{e}_{1}^{\intercal} = [1, 0, \dots, 0]^{\intercal} \in \R^{m} $ and
 $\beta =  \norm{\mathbf{\sigma}(0) + \mathbf{F}(t)}_{2}$. 
The residual of the approximation $\boldsymbol{\phi}_{m}(\tau)$ with respect
to the ODE system $\boldsymbol{\dot{\phi}}(t) = \mathbf{A} \boldsymbol{\phi}(t)$ is defined as:
 \begin{equation}
 \label{residual}
\mathbf{r}({m}, t) = \mathbf{A} \boldsymbol{\phi}_{m}(t) -  \boldsymbol{\dot{\phi}}_{m}(t) 
 \end{equation}
 
\noindent    
By replacing Eq. (\ref{decomp}) into Eq. (\ref{residual}), the residual norm relative
to the norm $\mathbf{\beta}$ is formed as:
 \begin{equation}
 \label{norm_residual}
\norm{\mathbf{r}({m}, t)} = | \beta\mathbf{h}_{m+1,m}\boldsymbol{\upsilon}_{m+1}\mathbf{e}_{m}^{\intercal}e^{t\mathbf{H}_{m}}\mathbf{e}_{1}  |
 \end{equation}
The proposed process is presented in Algorithm \ref{expsol_alg}.
As can be seen, in steps 3-19, it generates the EKS in $m$ iterations until the residual error of  Eq. (\ref{norm_residual}) is less than a tolerance error $\epsilon$, while in steps 9-14, it performs orthogonalization with respect to  $ \boldsymbol{\upsilon}_{1}, \dots, \boldsymbol{\upsilon}_{j}$ vectors. Finally, in step 22, it calculates the stress values at time $t$ by applying the generated EKS. 
The EKS has to be calculated  only once for transient analysis, after which it may be reused to evaluate the stress values at any time $t$.

\LinesNumbered

\begin{algorithm}[hbt!]
\SetAlgoLined
\DontPrintSemicolon
\KwIn{ matrix $\mathbf{A}_{E} \equiv \mathbf{G}^{-1}\mathbf{C}$, initial  vector $\mathbf{\sigma}(0)$, term $\mathbf{F}(t)$, desired order $m$, time $t$, budget error $\epsilon$ }
\KwOut{ vector of EM stress $\mathbf{\sigma}$ at time $t$ }

$\mathbf{B}_{E} = \mathbf{\sigma}(0) + \mathbf{F}(t)$\\
$\mathbf{V} = \texttt{qr}([\mathbf{B}_{E}, \mathbf{A}_{E}^{-1}\mathbf{B}_{E}])$\\

\For{$j=3,\dots,m$}{
    \eIf{ $ mod(j,2) = 0 $ }{
      $\mathbf{w} = \mathbf{A}_{E}^{-1}\boldsymbol{\upsilon}_{j-2}$\\
    }{
      $\mathbf{w} = \mathbf{A}_{E}\boldsymbol{\upsilon}_{j-2}$\\
    }
    \For{$i=1,\dots,j$}{
         $\mathbf{h}_{i,j} = \mathbf{w}^{T}\boldsymbol{\upsilon}_{i}$\\
         $\mathbf{w} = \mathbf{w} - \mathbf{h}_{i,j}\boldsymbol{\upsilon}_{i}$\\
    }
    $\mathbf{h}_{j+1,j} = \norm{\mathbf{w}}$ \\
    $\boldsymbol{\upsilon}_{j+1} = \frac{ \mathbf{w} }{\mathbf{h}_{j+1,j}}$\\
    \If{ $ \norm{\mathbf{r}({m}, t)} < \epsilon $ }{
      $ m = j$\\
      break
    }
}

$\mathbf{H}_{m} = \mathbf{H}(1:m,:) $ \\ 
$\mathbf{V}_{m} = \mathbf{V}(:,1:m) $  \\
$\mathbf{\sigma}(t) = \beta\mathbf{V}_me^{t\mathbf{H}_{m}}\mathbf{e}_{1} - \mathbf{F}(t)$

\caption{Proposed EKS-based EM stress~analysis }%
\label{expsol_alg}
\end{algorithm}
\normalsize

%% file: 4_Results.tex
\section{Experimental Evaluation}
\label{sec:experimentals}

In this Section, we present our experimental results that validate the reliability and scalability of the proposed methodology. 
First, in Section \ref{sec:accuracy_artificial}, our approach is compared to COMSOL v5.5, a Finite Element Method (FEM) based solver,  for an artificial multi-segment interconnect tree. 
Next, in Section \ref{sec:scalability}, 
we illustrate the scalability of our method for an interconnect tree with increasing number of via junctions (i.e., T junctions).
Finally,  in Section \ref{sec:OpenROAD}, we perform EM stress analysis on representative large-scale power grids to further evaluate the applicability of our method.
The characteristics of the Cu DD interconnects used in our simulations are the following: $Z^{*} = 1$, $e = 1.6\times10^{-19}$ $C$, $\rho = 2.25\times10^{-8}$ $\Omega m$, $\mathcal{B} = 28$ $GPa$, $\Omega = 1.18 \times 10^{-29}$ $m^3$, $D_{0} = 1.3\times10^{-9}$ $m^{2}/s$, $k = 1.38 \times 10^{-23}$ $J/K$, and $T = 378$ $K$.

The proposed approach was implemented in Matlab with default numerical packages, while the budget error $\epsilon$ of Algorithm~\ref{expsol_alg} was set to $10^{-3}$. 
Our experiments were performed on a  Windows workstation with 32 GB memory and a 4.7 GHz Intel Core i9-9900k processor with 16 threads. 

\subsection{Accuracy results on a seven-segment tree}
\label{sec:accuracy_artificial}
\begin{figure*}[!hbt]
    \begin{center} 	
    \includegraphics[width=1\linewidth]{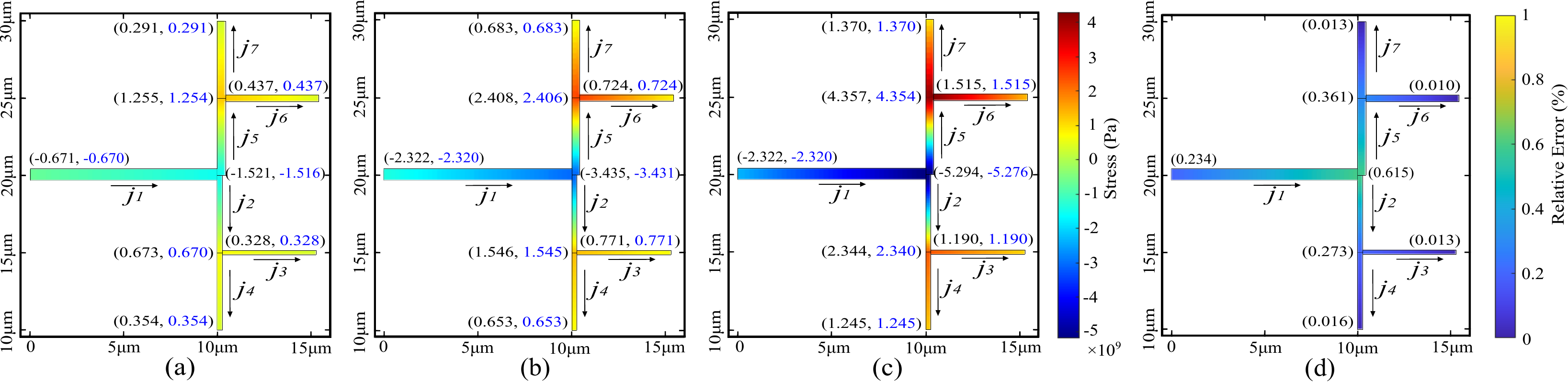}
    \end{center}
    \vspace{-8pt}
    \caption{(a)-(c): Comparison of EM stress at \boldsymbol{$t=5$}, \boldsymbol{$t=10$}, and \boldsymbol{$t=20$} years, respectively.
    The stress at each node is shown as a tuple, with our proposed solution in black and the COMSOL solution in blue text. (d): The relative error between  our proposed solution and the COMSOL solution at  \boldsymbol{$t=20$} years. 
    The widths of the tree are 
    \boldsymbol{$w_{1}=0.6$} \boldsymbol{$\mu m$},  \boldsymbol{$w_{2}=w_{3}=w_{4}=0.25$} \boldsymbol{$\mu m$}, \boldsymbol{$w_{5}=w_{6}=w_{7}=0.4$} \boldsymbol{$\mu m$}, while the lengths are shown in the figure. The current densities are \boldsymbol{$j_{1}=-2\times10^{-9}$} \boldsymbol{$A/m^{2}$}, \boldsymbol{$j_{2}=-3\times10^{-9}$} \boldsymbol{$A/m^{2}$}, \boldsymbol{$j_{5}=4\times10^{-9}$} \boldsymbol{$A/m^{2}$}, and \boldsymbol{$j_{3}=j_{4}=j_{6}=j_{7}=1\times10^{-9}$} \boldsymbol{$A/m^{2}$}. }
    \label{fig:accuracy_fig}
\end{figure*}

We constructed a seven-segment structure with three T junctions and assigned  different widths and current densities to each segment. 
Also, we performed transient simulation at $t=5$, $t=10$, and $t=20$ years, as shown in Figs. \ref{fig:accuracy_fig}(a)-(c), with the first colorbar depicting the spatial distribution  of the stress build-up  across the artificial tree. The tuples next to boundary or junction nodes are the values computed by our  method and COMSOL. In Fig. \ref{fig:accuracy_fig}(d), the relative error along the tree is found to be well below 0.5\% with a reduced order $m = 4$ after applying a discretization step $Dx = 2.5$~$\mu m$.
\begin{figure}[!t]
\vspace{-2mm}
    \begin{center} 	
    \includegraphics[width=0.8\linewidth,keepaspectratio]{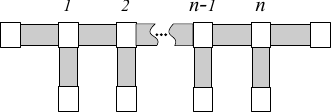}
    \end{center}
    \vspace{-8pt}
    \caption{ An interconnect with $\mathbf{n}$ successive  $\mathbf{T}$ juntions.}
    \label{fig:n_trees}
\end{figure}

\subsection{Scalability analysis}
\label{sec:scalability}

Table \ref{table_ntress} shows the runtime of the proposed method and COMSOL for increasing number of successive  T  junctions as shown in Fig. \ref{fig:n_trees}. Both methods calculate the hydrostatic stress at $t = 10$ years. Note that in an interconnect tree with  $n$ T junctions, there are $2n+1$ segments. Our method used a discretization step so as to formulate each segment into 20 points with the dimensions of horizontal segments being 10 $\mu$m in length and 0.4 $\mu$m in height, and the dimensions of vertical segments being 5 $\mu$m in length and 0.2 $\mu$m in height. 
In Table \ref{table_ntress}, $t_{\text{form}}$ is the runtime to build the LTI system including the runtime for singularity elimination; $t_{\text{exp\_init}}$ is the runtime to calculate the term $\mathbf{F}(t)$; and $t_{\text{exp\_sol}}$ is the total runtime of Algorithm \ref{expsol_alg}.
As shown in Table \ref{table_ntress}, the proposed method 
achieves up to 
268.74$\times$ runtime speedup over COMSOL.
Moreover, the calculation  of the LTI system, as well as the term $\mathbf{F}(t)$,  is the most time-consuming one of our proposed method. 
However, this calculation occurs only once prior to the transient analysis.

\begin{table}[!hbt]
\centering
\caption{
Runtime comparison of the proposed method and COMSOL for
increasing number of  T junctions}
\label{table_ntress}
\vspace{-7pt}
\resizebox{\columnwidth}{!}{%
\begin{tabular}{|c||cccc|c|c|}
\hline
\multirow{2}{*}{\textbf{n}} & \multicolumn{4}{c|}{\textbf{Proposed method (seconds)}}                                                              & \multirow{2}{*}{\textbf{\begin{tabular}[c]{@{}c@{}}COMSOL\\ (seconds)\end{tabular}}} & \multirow{2}{*}{\textbf{Speed-up}} \\ \cline{2-5}
                            & \multicolumn{1}{c|}{$t_{\text{form}}$} & \multicolumn{1}{c|}{$t_{\text{exp\_init}}$} & \multicolumn{1}{c|}{$t_{\text{exp\_sol}}$} & $t_{\text{all}}$ &                                                                                      &                                    \\ \hline\hline
100                         & \multicolumn{1}{c|}{0.013}      & \multicolumn{1}{c|}{0.002}           & \multicolumn{1}{c|}{0.015}          & 0.030      & 12                                                                                    & 267.91$\times$                      \\ \hline 
500                         & \multicolumn{1}{c|}{0.087}      & \multicolumn{1}{c|}{0.011}           & \multicolumn{1}{c|}{0.018}          & 0.116     & 28                                                                                    & 241.37$\times$                      \\ \hline
1000                        & \multicolumn{1}{c|}{0.246}      & \multicolumn{1}{c|}{0.032}           & \multicolumn{1}{c|}{0.017}          & 0.295     & 70                                                                                   & 237.28$\times$                      \\ \hline
2000                        & \multicolumn{1}{c|}{0.780}       & \multicolumn{1}{c|}{0.107}           & \multicolumn{1}{c|}{0.019}          & 0.906     & 232                                                                                   & 256.07$\times$                      \\ \hline
10000                       & \multicolumn{1}{c|}{16.341}     & \multicolumn{1}{c|}{2.427}           & \multicolumn{1}{c|}{0.022}          & 18.79     & 4824                                                                                  & 256.73$\times$                      \\ \hline
20000                       & \multicolumn{1}{c|}{35.753}     & \multicolumn{1}{c|}{10.382}          & \multicolumn{1}{c|}{0.027}          & 46.162   & 12406                                                                                 & 268.74$\times$                      \\ \hline
50000                       & \multicolumn{1}{c|}{75.674}    & \multicolumn{1}{c|}{39.467}          & \multicolumn{1}{c|}{0.049}          & 115.19    & 27193                                                                                & 236.07$\times$                      \\ \hline
\end{tabular}
}
\end{table}
\vspace{-3mm}
\subsection{Analysis on OpenROAD power grid designs}
\label{sec:OpenROAD}
In order to perform analysis on large-scale power grids, we employed several OpenROAD
circuits designed using a commercial 12 nm FinFET, commercial
28 nm FDSOI, and Nangate 45 nm  technologies. 
More specifically, these benchmarks are built as mesh-like orthogonal structures, on which
we employ the BFS traversal algorithm to identify the boundary segments as well as the via segments and decompose the meshes into multi-segment trees.
The fact that COMSOL is inefficient  for very large circuits, led us to analyze only the tree with the largest number of segments and vias (denoted as the "largest tree") from each benchmark,  since it indicates the worst-case scenario in terms of runtime and accuracy. 
For this experiment, we  performed transient analysis at $t=20$ years.

The  results for the largest tree of each OpenROAD benchmark are summarized in Table \ref{table_openroad}, where \textit{\#Segments} represents the total number of horizontal and vertical (via) segments per tree and \textit{\#T junction} describes the number of T junctions per tree.
As can be seen, our method maintains a relative error lower than 1\%, even for the largest circuits, error reduction percentage of at least 99.5\%, while achieving a 252.03$\times$ average  runtime speedup over COMSOL.


%% file: 5_Conclusions.tex
\section{Conclusions}
\label{sec:conclusions}
In this paper, we proposed a fast semi-analytical approach based on the matrix exponential for the computation of EM stress at discrete spatial points of interconnect trees. The main idea of our approach is to apply the EKS to accelerate  the computation of the matrix exponential and enable the efficient analysis of large-scale models. Experimental results on artificial interconnect trees and representative OpenROAD power grids indicate that our proposed approach is three orders of magnitude faster than  the industrial FEM-based solver COMSOL while providing similar~accuracy.

%% file: acknowledgement.tex
\section*{Acknowledgments}
This research has been co-financed by the European Regional Development Fund and Greek national funds via the Operational Program "Competitiveness, Entrepreneurship and Innovation",~under the call "RESEARCH-CREATE-INNOVATE" (project code: T2EDK-00609).
\vspace{2mm}
\begin{table*}[!hbt]
\centering
\caption{Runtime and accuracy comparison between the proposed method and COMSOL for
large-scale OpenROAD power~grids}
\label{table_openroad}
\vspace{-6pt}
\begin{tabular}{|c|c||ccccc||ccc||c|}
\hline
\multirow{4}{*}{\textbf{Tech.}} & \multirow{4}{*}{\textbf{Design}} & \multicolumn{5}{c||}{\textbf{Largest tree}}                                                                                                                                                                                                                                                                                                                                                                                                                                        & \multicolumn{3}{c||}{\textbf{Runtime}}                                                                                                                                                                                                                                   & \textbf{Accuracy}                                                                            \\ \cline{3-11} 
                               &                                  & \multicolumn{1}{c|}{\multirow{3}{*}{\textbf{\#Segments}}} & \multicolumn{1}{c|}{\multirow{3}{*}{\textbf{\begin{tabular}[c]{@{}c@{}}\#T \\ junctions\end{tabular}}}} & \multicolumn{1}{c|}{\multirow{3}{*}{\textbf{\begin{tabular}[c]{@{}c@{}}Discr.\\ step\\ ($\boldsymbol{\mu m}$)\end{tabular}}}} & \multicolumn{1}{c|}{\multirow{3}{*}{\textbf{\begin{tabular}[c]{@{}c@{}}Initial\\ size\end{tabular}}}} & \multirow{3}{*}{\textbf{\begin{tabular}[c]{@{}c@{}}Reduced\\ size\end{tabular}}} & \multicolumn{1}{c|}{\multirow{3}{*}{\textbf{\begin{tabular}[c]{@{}c@{}}Proposed\\ method  \\ (seconds)\end{tabular}}}} & \multicolumn{1}{c|}{\multirow{3}{*}{\textbf{\begin{tabular}[c]{@{}c@{}}COMSOL\\ (seconds)\end{tabular}}}} & \multirow{3}{*}{\textbf{Speed-up}} & \multirow{3}{*}{\textbf{\begin{tabular}[c]{@{}c@{}}Relative\\ error\end{tabular}}} \\ 
                               &                                  & \multicolumn{1}{c|}{}                                     & \multicolumn{1}{c|}{}                                                                                    & \multicolumn{1}{c|}{}                                                                                           & \multicolumn{1}{c|}{}                                                                                 &                                                                                  & \multicolumn{1}{c|}{}                                                                                                  & \multicolumn{1}{c|}{}                                                                                     &                                    &                                                                                              \\
                               &                                  & \multicolumn{1}{c|}{}                                     & \multicolumn{1}{c|}{}                                                                                    & \multicolumn{1}{c|}{}                                                                                           & \multicolumn{1}{c|}{}                                                                                 &                                                                                  & \multicolumn{1}{c|}{}                                                                                                  & \multicolumn{1}{c|}{}                                                                                     &                                    &                                                                                              \\ \hline
                               \hline
\multirow{5}{*}{45 nm}         & dynamic                  & \multicolumn{1}{c|}{640}                                  & \multicolumn{1}{c|}{77}                                                                                  & \multicolumn{1}{c|}{0.25}                                                                                       & \multicolumn{1}{c|}{12801}                                                                            & 5                                                                                & \multicolumn{1}{c|}{0.063}                                                                                            & \multicolumn{1}{c|}{12}                                                                                    & 190.47$\times$                      & 0.14\%                                                                                       \\ \cline{2-11} 
                               & ibex                             & \multicolumn{1}{c|}{1096}                                 & \multicolumn{1}{c|}{107}                                                                                 & \multicolumn{1}{c|}{0.25}                                                                                       & \multicolumn{1}{c|}{21920}                                                                            & 12                                                                               & \multicolumn{1}{c|}{0.124}                                                                                           & \multicolumn{1}{c|}{21}                                                                                    & 169.35$\times$                      & 0.15\%                                                                                      \\ \cline{2-11} 
                               & aes                              & \multicolumn{1}{c|}{1327}                                 & \multicolumn{1}{c|}{82}                                                                                  & \multicolumn{1}{c|}{0.25}                                                                                       & \multicolumn{1}{c|}{26541}                                                                            & 11                                                                               & \multicolumn{1}{c|}{0.162}                                                                                           & \multicolumn{1}{c|}{26}                                                                                    & 160.49$\times$                      & 0.29\%                                                                                      \\ \cline{2-11} 
                               & jpeg                             & \multicolumn{1}{c|}{3337}                                 & \multicolumn{1}{c|}{184}                                                                                 & \multicolumn{1}{c|}{0.25}                                                                                       & \multicolumn{1}{c|}{66741}                                                                            & 13                                                                               & \multicolumn{1}{c|}{0.716}                                                                                           & \multicolumn{1}{c|}{111}                                                                                   & 155.02$\times$                      & 0.51\%                                                                                      \\ \cline{2-11} 
                               & swerv                            & \multicolumn{1}{c|}{5226}                                 & \multicolumn{1}{c|}{27}                                                                                  & \multicolumn{1}{c|}{0.25}                                                                                       & \multicolumn{1}{c|}{104521}                                                                           & 17                                                                               & \multicolumn{1}{c|}{1.640}                                                                                           & \multicolumn{1}{c|}{254}                                                                                   & 154.87$\times$                      & 0.59\%                                                                                      \\ \hline \hline
\multirow{3}{*}{28 nm}         & gcd                              & \multicolumn{1}{c|}{25}                                   & \multicolumn{1}{c|}{7}                                                                                   & \multicolumn{1}{c|}{1}                                                                                          & \multicolumn{1}{c|}{1001}                                                                             & 6                                                                                & \multicolumn{1}{c|}{0.018}                                                                                           & \multicolumn{1}{c|}{4}                                                                                    & 222.22$\times$                      & 0.37\%                                                                                      \\ \cline{2-11} 
                               & aes                              & \multicolumn{1}{c|}{421}                                  & \multicolumn{1}{c|}{159}                                                                                 & \multicolumn{1}{c|}{1}                                                                                          & \multicolumn{1}{c|}{16841}                                                                            & 8                                                                                & \multicolumn{1}{c|}{0.043}                                                                                           & \multicolumn{1}{c|}{10}                                                                                    & 232.55$\times$                      & 0.55\%                                                                                      \\ \cline{2-11} 
                               & jpeg                             & \multicolumn{1}{c|}{2285}                                 & \multicolumn{1}{c|}{320}                                                                                 & \multicolumn{1}{c|}{1}                                                                                          & \multicolumn{1}{c|}{91401}                                                                            & 14                                                                               & \multicolumn{1}{c|}{0.377}                                                                                           & \multicolumn{1}{c|}{56}                                                                                    & 148.54$\times$                      & 0.63\%                                                                                      \\ \hline \hline
\multirow{5}{*}{12 nm}         & gcd                              & \multicolumn{1}{c|}{276}                                  & \multicolumn{1}{c|}{73}                                                                                  & \multicolumn{1}{c|}{0.1}                                                                                        & \multicolumn{1}{c|}{5520}                                                                             & 4                                                                                & \multicolumn{1}{c|}{0.034}                                                                                           & \multicolumn{1}{c|}{8}                                                                                    & 235.29$\times$                      & 0.64\%                                                                                      \\ \cline{2-11} 
                               & ibex                             & \multicolumn{1}{c|}{1486}                                 & \multicolumn{1}{c|}{165}                                                                                 & \multicolumn{1}{c|}{0.1}                                                                                        & \multicolumn{1}{c|}{29721}                                                                            & 9                                                                                & \multicolumn{1}{c|}{0.184}                                                                                           & \multicolumn{1}{c|}{44}                                                                                    & 239.13$\times$                      & 0.53\%                                                                                      \\ \cline{2-11} 
                               & jpeg                             & \multicolumn{1}{c|}{5702}                                 & \multicolumn{1}{c|}{326}                                                                                 & \multicolumn{1}{c|}{0.1}                                                                                        & \multicolumn{1}{c|}{114041}                                                                           & 15                                                                               & \multicolumn{1}{c|}{1.936}                                                                                           & \multicolumn{1}{c|}{505}                                                                                   & 260.84$\times$                      & 0.54\%                                                                                      \\ \cline{2-11} 
                               & dynamic                    & \multicolumn{1}{c|}{10304}                                & \multicolumn{1}{c|}{437}                                                                                 & \multicolumn{1}{c|}{0.1}                                                                                        & \multicolumn{1}{c|}{206081}                                                                           & 20                                                                               & \multicolumn{1}{c|}{5.860}                                                                                           & \multicolumn{1}{c|}{1712}                                                                                  & 292.15$\times$                      & 0.97\%                                                                                      \\ \cline{2-11} 
                               & aes                              & \multicolumn{1}{c|}{13148}                                & \multicolumn{1}{c|}{453}                                                                                 & \multicolumn{1}{c|}{0.1}                                                                                        & \multicolumn{1}{c|}{262961}                                                                           & 23                                                                               & \multicolumn{1}{c|}{10.768}                                                                                          & \multicolumn{1}{c|}{2874}                                                                                  & 266.90$\times$                      & 0.82\%                                                                                      \\ \hline
\end{tabular}
\end{table*}